\documentclass[showpacs,preprint,aps]{revtex4}%
\usepackage{amsfonts}
\usepackage{amsmath}
\usepackage{amssymb}
\usepackage{epsfig}
\usepackage{graphicx}%
\ifx\pdfoutput\relax\let\pdfoutput=\undefined\fi
\newcount\msipdfoutput
\ifx\pdfoutput\undefined\else
\ifcase\pdfoutput\else
\ifx\paperwidth\undefined\else
\ifdim\paperheight=0pt\relax\else\pdfpageheight\paperheight\fi
\ifdim\paperwidth=0pt\relax\else\pdfpagewidth\paperwidth\fi
\fi\fi\fi
\begin{document}
\title{Stability of Phase-modulated Quantum Key Distribution System}
\author{Zheng-Fu Han}
\email{zfhan@ustc.edu.cn}
\author{Xiao-Fan Mo}
\author{You-Zhen Gui}
\author{Guang-Can Guo}
\email{gcguo@ustc.edu.cn}
\affiliation{Key Laboratory of Quantum Information, University of Science and Technology of
China (CAS), Hefei, Anhui 230026, People's Republic of China.}

\begin{abstract}
Phase drift and random fluctuation of interference visibility in double
unbalanced M-Z QKD system are observed and distinguished. It has been found
that the interference visibilities are influenced deeply by the disturbance of
transmission fiber. Theory analysis shows that the fluctuation is derived from
the envioronmental disturbance on polarization characteristic of fiber,
especially including transmission fiber. Finally, stability conditions of
one-way anti-disturbed M-Z QKD system are given out, which provides a
theoretical guide in pragmatic anti-disturbed QKD.

\end{abstract}

\maketitle

Quantum cryptography (QC) or quantum cryptographic key distribution (QKD) has
advanced for twenty years since Bennett and Brassard proposed the idea in
$1984$ \cite{Bennett1}. It has approached practicability to date, and is
applied in the field of information security \cite{Stucki,www}. Many protocols
for QC are proposed up to the present, including BB84 \cite{Bennett1}, B92
\cite{Bennett2} and improved protocols \cite{Bruss,Bourennane} based on single
photons, and EPR protocols \cite{Ekert} based on entangled photon pairs. It
has been proved that BB84 protocol and EPR protocol are consistent essence
\cite{Bennett4}. Considering current experimental technique, single-photon
protocols are more practical for commercial QKD.

A general challenge for QKD is to choose a coding scheme free of disturbance
by the dominant noise sources. In the case of single-photon protocols, quantum
information is encoded by the states of single photons that travel from Alice
to Bob, and two kinds of schemes are utilized: polarization coding and phase
coding. Polarization coding is adopted in free-space QKD because atmospheric
density fluctuations only reduce the collection efficiency but don't degrade
the contrast between the polarization basis states. C. Kurtsiefer \textit{et
al.} have realized the secure exchange of keys over a free space path of
$23.4$ kilometres between Zugspitze and karwendespitze of the Alps
\cite{Kurtsiefer}. This marks a pioneering step towards accomplishing key
exchange with a near-Earth orbiting satellite and hence a global
key-distribution system \cite{Rarity}. On the other hand, optical fibres are
appropriate for communication on the ground without influence of weather and
atmospheric pollution. However, it is a calamity to use polarization as basis
for encoding quantum information due to significant birefringence. Hence,
phase coding is widely used in fibre-optic QKD-prototype systems.

The most typical fiber QKD prototype was designed in Ref. \cite{Bennett3}. In
that prototype, two photon pulses pass through the same transmission fiber
(quantum channel), and the same disturbance, from transmission processes, is
expected to be counteracted on Bob's side. But in fact, this prototype gives
us a bad systemic stability. In order to improve the stability, two groups
\cite{Muller,Bethune}\ developed a new prototype independently, in which the
two pulses transmit a round trip with a Faraday reflection in mid-course.
Unfortunately, this prototype leaves a chance to Eavesdroppers in the key
exchange. Eve can send Trojan-horse photons to tail signal photons thus pass
in and out of Bob's secure office, collecting Bob's coded information without
being discovered \cite{Boileau}. Furthermore, any go-and-return QKD protocol
\cite{Nishioka} is essentially unsecured, and true secure QKD should be based
on photons propagating along one way. Therefore, for\ practical use of QKD, it
is very important to investigate the stability of double M-Z interferometers.
This paper investigates the stability in experiments and establishes a
theoretical model. Two kinds of perturbation motion, i.e., phase drift and
random fluctuation of interference visibility, are observed and discussed.
Formerly, it was ever believed that the dominant disturbance arises from
different environments of interferometers on Alice and Bob's sides, and the
disturbance in quantum channel is equivalent to both pulses (with $%
%TCIMACRO{\unit{ns}}%
%BeginExpansion
\operatorname{ns}%
%EndExpansion
$-scale time interval) which can be counteracted in the final interference.
But our experiments show that interference visibility depends intensively on
length of transmission fiber ($L$) and systemic stability is also influenced
by the way disturbance. Theoretical analysis, here, deduces the conditions for
systemic stabilization of double unbalanced M-Z interferometers. It will offer
important reference for stable one-way QKD systems.

Typical QKD prototype is described in Fig. 1. Two uniform unbalanced M-Z
interferometers were built with common single mode fiber (SMF-28), which is
respectively the coder of Alice or Bob. Quantum channel between Alice and Bob
and beamsplitters ($50/50$) are also made of the same fiber. Experimental
result is depicted in Fig. 2. Fig. 2a shows the received optical power of
single detector D1 for $L=2%
%TCIMACRO{\unit{m}}%
%BeginExpansion
\operatorname{m}%
%EndExpansion
$ without any phase modulation. Apparently, the fluctuation is derived from
phase drift, which is relevant to the environment of interferometers. The main
reason is that environmental temperature has a small fluctuation, and is also
influenced by other perturbation motion, for instance, fiber vibration. Fig.
2b describes the interference fringe of the QKD system when $L=75$ $%
%TCIMACRO{\unit{km}}%
%BeginExpansion
\operatorname{km}%
%EndExpansion
$, and here the phase modulator is driven by long-periods saw-tooth wave. The
fluctuation of envelope of interference fringe exceeds $50\%$ and shows full
randomness within six hours. The random fluctuation should be closely
associated with the environment of the system.

Summarizing the results of Fig. 2, systemic instability resulted from
environmental fluctuation behave as phase drift and random fluctuation of
interference visibilities. The former leads to work-point destabilization for
QKD\ system and can be corrected by instantaneous calibration. But the latter
results in qubit-error-rate rising, which can not be artificially controlled
due to randomness of environmental fluctuation. In principle, rigorous
anti-disturbance methods can be used in Alice and Bob's secure offices, so the
effect of the disturbance can decrease to as weaker as possible. However,
coupling between quantum channel and its environmental disturbance is
uncontrolled. Therefore, disturbance from environment of transmission fiber is
discussed as follows.

Fig. 3 shows practical interference visibilities for different transmission
fiber length $L$. For $L=0$, the fluctuation of the visibility does not exceed
$5\%$ within $280$ hours. For $L=25,50,75$ $%
%TCIMACRO{\unit{km}}%
%BeginExpansion
\operatorname{km}%
%EndExpansion
$, this fluctuation is more than $80\%$. It is most interesting, that the
longer the transmission fiber is, the faster\ the visibilities fluctuate.
These results indicate that visibilities are affirmatively influenced by way
disturbance. Additional consideration is that, temperature fluctuation only
leads to phase drift but not interference visibilities fluctuation. Therefore,
fluctuation of interference visibilities should arise from polarization
birefringence in fiber induced by perturbation motions, such as fiber bend and
distortion. In order to explore the relationships between interference
visibility and polarization characteristic in M-Z systems, a simple but
reasonable theoretical model is built here.

For the sake of clarity and concision, we assume that there is no fiber
nonlinearity and no polarization-dependent loss, and that the usual loss terms
of the fiber have been factored out so that we can deal with unitary Jones
matrix \cite{Jones} to describe polarization-transport character of each part
in the system (interferometers of Alice and Bob, transmission fiber).
$A_{1},A_{2},C,B_{1},B_{2}$ are respectively Jones matrixes to each part of
fiber in the QKD apparatus (see Fig. 1). Two paths leading to single-photon
interference on $BS_{4}$ are: $P_{1}\left(  A_{1}\rightarrow C\rightarrow
B_{2}\rightarrow PM_{B}\right)  $ and $P_{2}\left(  A_{2}\rightarrow
PM_{A}\rightarrow C\rightarrow B_{1}\right)  $. Here $PM_{A}$ and $PM_{B}$ are
two phase modulators. So single-photon optical transformation matrixes through
two paths are respectively described by:%
\begin{align}
B_{2}e^{i\beta_{2}}\cdot e^{i\varphi_{B}}\cdot Ce^{i\phi}\cdot A_{1}%
e^{i\alpha_{1}}  &  =B_{2}CA_{1}e^{i\left(  \alpha_{1}+\beta_{2}+\phi
+\varphi_{B}\right)  },\\
B_{1}e^{i\beta_{1}}\cdot Ce^{i\phi}\cdot A_{2}e^{i\alpha_{2}}\cdot
e^{i\varphi_{A}}  &  =B_{1}CA_{2}e^{i\left(  \alpha_{2}+\beta_{1}+\phi
+\varphi_{A}\right)  },
\end{align}
where $\alpha_{i},\beta_{i}$ are common phases through $i$th fiber of Alice or
Bob and $\varphi_{A}\left(  \varphi_{B}\right)  $ is modulated phase of
$PM_{A}\left(  PM_{B}\right)  $.

Supposing input Jones vector of field \cite{Jones} is $E_{in}$, and each
effective Jones vector arriving at $BS_{4}$ is $E_{in}/4$. Hence input power
$I_{in}=\left\vert E_{in}\right\vert ^{2}$ and the output Jones vector can be
written:%
\begin{align}
E_{out}  &  =\left[  B_{2}CA_{1}e^{i\left(  \alpha_{1}+\beta_{2}+\phi
+\varphi_{B}\right)  }\right. \nonumber\\
&  \left.  +B_{1}CA_{2}e^{i\left(  \alpha_{2}+\beta_{1}+\phi+\varphi
_{A}\right)  }\right]  \frac{E_{in}}{4}.
\end{align}
On the consideration that $A_{i},B_{i},C$ are unitary, output power can be
expressed as:%
\begin{align}
P_{out}  &  =E_{out}^{+}\cdot E_{out}\nonumber\\
&  =\frac{P_{in}}{8}+\frac{1}{16}E_{in}^{+}\left[  A_{1}^{+}C^{+}B_{2}%
^{+}B_{1}CA_{2}e^{-i\left(  \Delta\alpha+\Delta\beta+\Delta\varphi\right)
}\right. \nonumber\\
&  \left.  +A_{2}^{+}C^{+}B_{1}^{+}B_{2}CA_{1}e^{i\left(  \Delta\alpha
+\Delta\beta+\Delta\varphi\right)  }\right]  E_{in},
\end{align}
where $\Delta\alpha=\alpha_{1}-\alpha_{2},\Delta\beta=\beta_{1}-\beta
_{2},\Delta\varphi=\varphi_{B}-\varphi_{A}$. Note that the polarized
fluctuation, is random, and consequently the matrixes $A_{i},B_{i},C$ are not
independent on time. Obviously, $P_{out}$ is a function of $A_{i}%
,B_{i},C,\Delta\alpha,\Delta\beta,\Delta\varphi$, which means interferential
output is dependent not only on both M-Z interferometers but also on
transmission fiber. In fact, disturbance from M-Z interferometers and
transmission fiber are not independent, which supports our experimental results.

In Eq. 4, if we make $B_{1}^{+}B_{2}=I$ on Bob's side, then:%
\begin{align}
P_{out}  &  =\frac{P_{in}}{8}+\frac{1}{16}E_{in}^{+}\left[  A_{1}^{+}%
A_{2}e^{-i\left(  \Delta\alpha+\Delta\beta+\Delta\varphi\right)  }\right.
\nonumber\\
&  \left.  +A_{2}^{+}A_{1}e^{i\left(  \Delta\alpha+\Delta\beta+\Delta
\varphi\right)  }\right]  E_{in},
\end{align}
where we have considered $C$ is unitary and thus $C^{+}C=I$. Now $P_{out}$ no
longer depends on transmission fiber, i.e., $P_{out}$ is independent on\ any
polarized disturbance in transmission\ fiber. Furthermore, if $A_{1}^{+}%
A_{2}=I$ on Alice's side, Eq. 5 is simplified as%
\begin{equation}
P_{out}=P_{in}\frac{1+\cos\left(  \Delta\alpha+\Delta\beta+\Delta
\varphi\right)  }{8}.
\end{equation}
This means that interferential output power $P_{out}$ does not rely on any
polarized perturbation motion in whole QKD system. In the ideal case,
$\Delta\alpha,\Delta\beta$ are invariable, and hence interference fringes are
only modulated by $\Delta\varphi$ brought by both phase modulators. However,
in the actual case, the fluctuation of environmental temperature will bring
some drift of $\Delta\alpha,\Delta\beta$, corresponding to Fig. 2a. In fact,
the fluctuation of environmental temperature is so slow that one can calibrate
it by instantaneous compensation.

Aforementioned anti-disturbance conditions $A_{1}^{+}A_{2}=I,B_{1}^{+}B_{2}=I$
can be equivalently written as:%
\begin{equation}
A_{1}=A_{2},B_{1}=B_{2},
\end{equation}
because of unitarity of matrixes $A_{i},B_{i}$. These mean that if we can
assure uniform polarization character for two arms of each interferometer,
then the QKD\ system resists polarized disturbance not only from transmission
fiber (quantum channel) but also from the interferometers. A typical prototype
to satisfy upper conditions is that the inner light paths of interferometers
are polarization-maintaining. In this case, the Jones matrixes are%
\begin{equation}
A_{1}=A_{2}=I,B_{1}=B_{2}=I.
\end{equation}
Obviously, Eq. 8 is a special case to Eq. 7, and it points out that each arm
must maintain input photon's polarization. In experiment, free space or
polarization-maintaining fibers of interferometers can fulfill Eq. 8.

We have observed and distinguished two kinds of unstable phenomena in double
unbalanced M-Z QKD system, i.e., phase drift and random fluctuation of
interference visibilities. We have found that the interference visibilities
can be influenced deeply by the disturbance of transmission fiber. Through
theoretic analysis, it is pointed out that the influences are derived from the
environmental disturbance on polarization characteristic of fiber, and give
out the stability condition of anti-disturbance M-Z QKD system with one-way
photons transmission. The conditions contribute to the researches on pragmatic
no round trip anti-disturbed QKD system.

We thank associate Prof. Bing Zhu, for his helpful advice. This work was
funded by National Fundamental Research Program of China (2001CB309301), also
by National Natural Science Foundation of china (60121503) and the Innovation
Funds of Chinese Academy of Sciences.

\begin{figure}[ptb]
\includegraphics[scale=0.5]{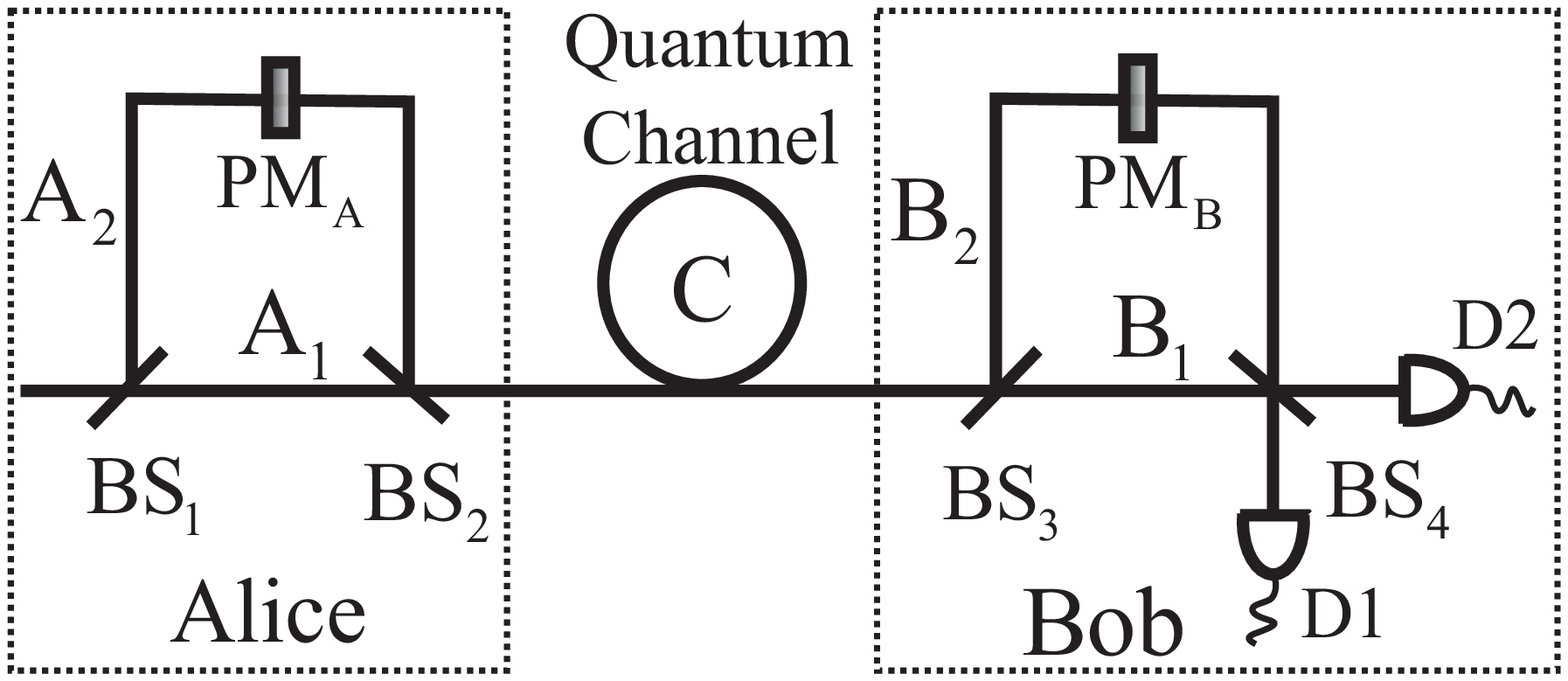}\caption{QKD system based on double
unbalanced Mach-Zehnder interferometers.}%
\end{figure}

\begin{figure}[ptb]
\includegraphics[scale=1.0]{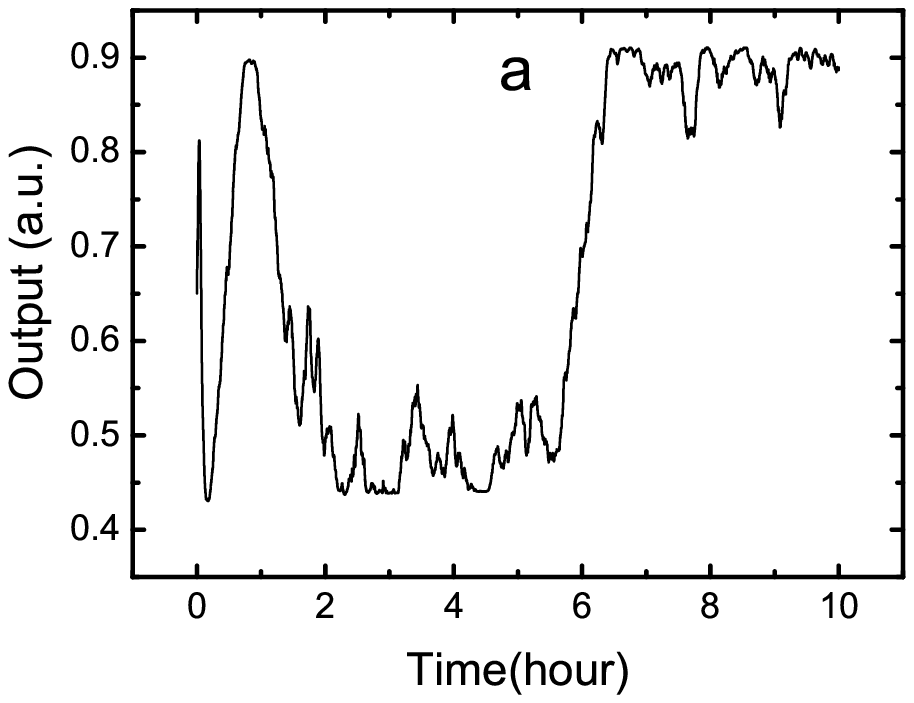}\end{figure}

\begin{figure}[ptb]
\includegraphics[scale=1.0]{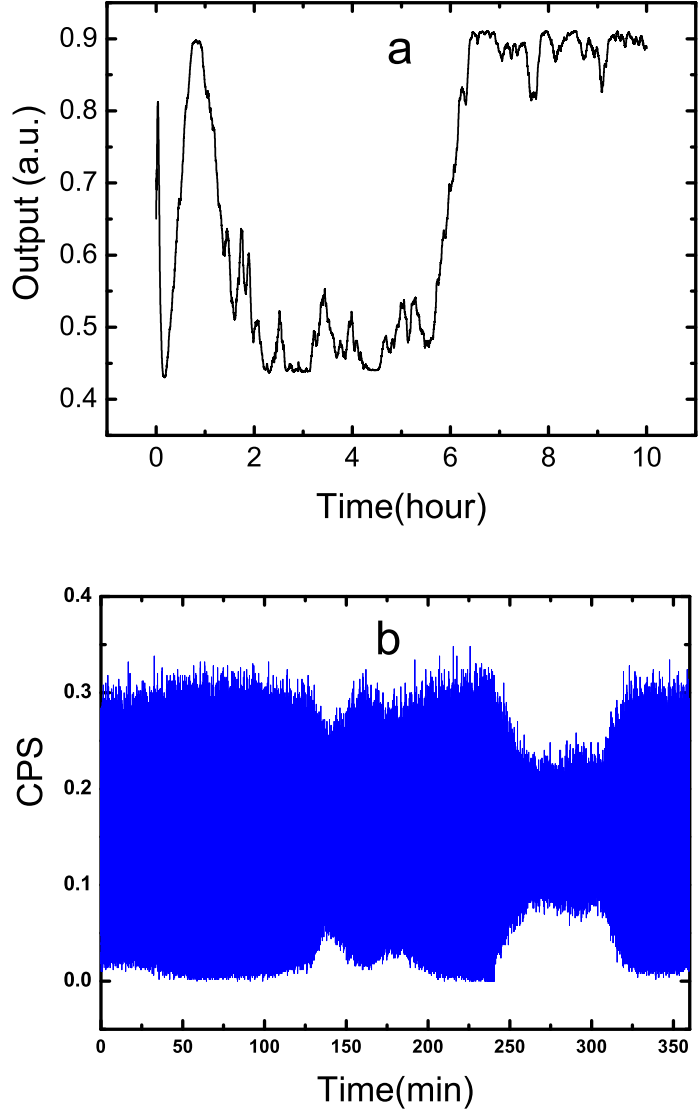}\caption{(a) Output power of single
detector D1 without any phase modulation, here $L=2\operatorname{m}$, and
shows phase excursion for different time. (b) Interference fringe using the
QKD system for different time, here $L=75$ $\operatorname{km}$. Phase
modulator is controlled by long-periods saw-tooth wave.}%
\end{figure}

\begin{figure}[ptb]
\includegraphics[scale=1.0]{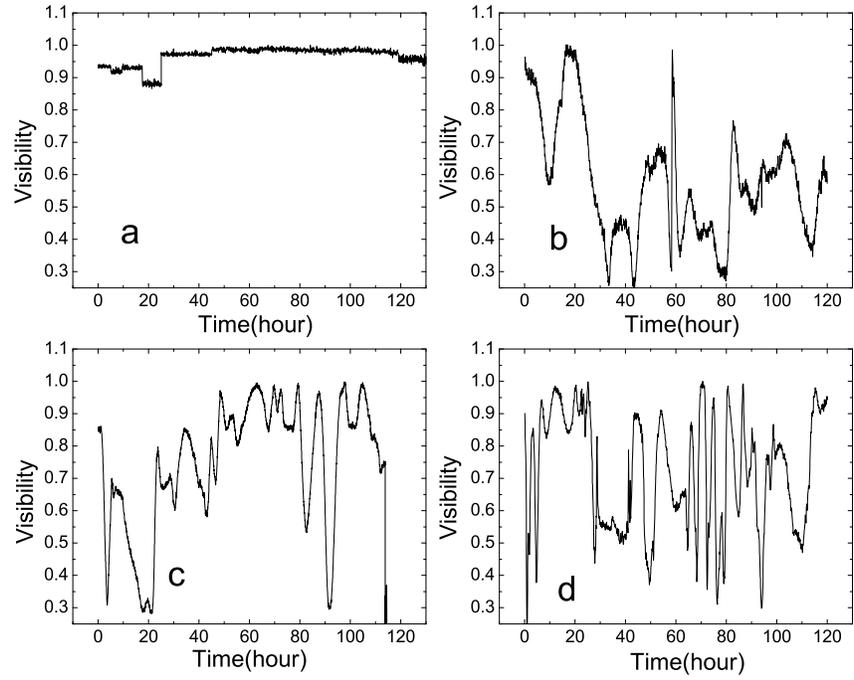}\caption{Interference visibilities based
on double unbalanced M-Z interferometers at different time. a: $L=0$ ; b:
$L=25$ $\operatorname{km}$; c: $L=55$ $\operatorname{km}$; d: $L=75$
$\operatorname{km}$.}%
\end{figure}

\end{document}